\documentclass[%
 reprint,
 amsmath,amssymb,
 aps,
 pra,
]{revtex4-2}

\usepackage{graphicx}
\usepackage{dcolumn}
\usepackage{bm}
\usepackage{amsfonts}
\usepackage{amsmath}
\usepackage{float}
\usepackage{caption}
\usepackage{color}
\usepackage{comment}
\usepackage{soul}

\usepackage{graphicx,subcaption}

\begin{document}

\preprint{APS/123-QED}

\title{Topology-Inspired Interfaces in Non-Local Mass-Spring Lattices}

\author{R.~G. Edge, T.~A. Starkey, and G.~J. Chaplain}

\affiliation{Centre for Metamaterial Research and Innovation, Department of Physics and Astronomy, University of Exeter, Exeter EX4 4QL, United Kingdom}

\date{\today}

\begin{abstract}
   Non-local couplings offer a means of engineering the dynamics and topology of discrete lattices beyond the constraints of nearest-neighbour (NN) interactions. Here, we extend the mass-spring Su–Schrieffer–Heeger (SSH) model to incorporate beyond-nearest-neighbour (BNN) connections that preserve the chiral symmetry of the diatomic unit cell, investigating how their range and strength shape the bulk response. We show that the accessible winding numbers are bounded by the lattice connectivity, while the spring stiffnesses select the specific topological phase. We further investigate the challenges posed by non-local interfaces, where the extended connectivity inherently introduces local boundary perturbations. The dynamic freedom afforded by BNN couplings is exploited by examining interfaces between dynamically and topologically dissimilar bulk media, revealing a band-gap-induced asymmetry in the localisation of interface states. We show that the interfacial configuration provides a means of controlling the existence and character of localised interface modes, with changes to the interface enabling the tuning of mode frequencies, localisation and hybridisation with the bulk bands.
\end{abstract}

\maketitle
\section{Introduction}

Topology characterises the global properties of a system that are preserved under continuous deformations, independent of specific geometric details; in physical systems, these properties can give rise to boundary states that are robust against perturbations \cite{Kane2013, Hasan2010}. Invariants such as the Berry phase and winding number provide a quantitative framework for characterising topological behaviour \cite{Horsley2024}, enabling the prediction of emergent phenomena such as protected edge modes and interface states. By leveraging these global principles, systems can be designed to reliably remain resilient to imperfections \cite{Wagner2016, Bernard2026, Paz2026}, a crucial advantage in real-world applications where fabrication and operational defects are inevitable. This robustness against disorder unlocks a wide array of practical applications and advanced functionalities. Wave guiding has been realised at the boundaries between distinct topological media \cite{Miniaci2021, Poggetto2024}, and robust topological effects have been harnessed in energy and vibration control \cite{Chen2024, Chaplain2020, Chaplain2020}. Devices such as sensors and insulators have also been shown to benefit from these topological phenomena \cite{Kumar2022, Hasan2010}.

An elementary introduction to topological concepts is often provided through the one-dimensional Su-Schrieffer-Heeger (SSH) model, originally developed to describe the electronic properties of Polyacetylene \cite{Su1979}. The SSH model has since become a quintessential example of a one-dimensional topological insulator, illustrating key concepts such as topological phase transitions, edge states, and chiral symmetry protection \cite{Obana2019}. Its simplicity and versatility have allowed the SSH model to be widely adopted in mechanical, photonic, and cold-atom systems to study wave phenomena and topological effects in both discrete and continuous settings \cite{Diakonos2025, Rontgen2024, Lix2018}. In its canonical form, however, the model is restricted to nearest-neighbour (NN) couplings.

The study of wave dispersion in periodic structures is often traced back to the foundational work of Brillouin \cite{Brillouin1953}, with more recent research demonstrating how beyond-nearest-neighbour (BNN) connections can give rise to exotic dispersion characteristics and wave propagation phenomena \cite{Chen2021, Martinez, EdgeEB2025}. These extended interactions offer a versatile platform for tailoring band structures, controlling group velocities, and exploring novel dynamic effects in lattice systems \cite{Chen2025}. The flexibility afforded by BNN couplings not only enriches the fundamental physics of periodic lattices but also provides a powerful toolkit for designing bespoke dynamic responses \cite{Kazemi2023, EdgeMass2025, Paul2025}, making them a highly promising platform for advancing wave-based technologies. Tuning both the spatial extent and strength of the connections has enabled a range of applications, however, despite the capabilities of BNNs being well-documented, rarely are the topological implications considered.

The inherent complexity of non-local lattices often leads to topology-agnostic studies, which focus instead on other aspects of the system, such as dispersion engineering and dynamic wave phenomena. Despite this, some studies have investigated the emergence of edge states in non-local structures, probing the interplay between extended interactions and topological effects. In condensed-matter, and superconducting systems, this has been examined most thoroughly through Kitaev and Ising chains \cite{Vodola2014,Baghran2024, Berganza2015, Ren2026}. Long-range connections have been shown to modify phase boundaries, and produce new topological phases, however in some settings a weakening of the bulk-boundary correspondence is observed, with the correspondence breaking down entirely in the strong long-range limit, whereby the bulk invariant alone no longer determines the number of edge modes \cite{Shi2025, Lepori2017}. Studies of non-locality in classical wave systems remain comparatively limited and fragmented. Non-local discrete mass–spring SSH chains have been used to investigate in-gap edge and domain-wall states \cite{Li2018, Rajabpoor2024, Rajabpoor2023}, while long-range hopping in plasmonic chains have been shown to modify edge-mode behaviour \cite{Pocock2019}. Feedback-based control has provided a means to engineer long-range, non-reciprocal couplings in active acoustic and mechanical metamaterials, offering new routes to topological phenomena \cite{Sirota2020, Braghini2022, SWang2023, Ruzzene2020}. More recently, gapless topology has emerged as a framework for understanding topological features at criticality, where the bulk band gap closes and conventional topological invariants become ill-defined. Remarkably, protected boundary modes can persist even in this gapless regime \cite{Cheng2026, Lin2026}. However, these studies are typically confined to a fixed, single order of non-locality, focusing mainly on edge states, leaving the natural extension to interface states largely unexplored.

In this work, we extend the principles of the discrete mass-spring SSH model to incorporate non-local connections; retaining an underlying diatomic structure and preserving chiral symmetry protection in the bulk material. We characterise the topology of the bulk medium, evaluating how both the order and strength of the connections modifies topological invariants, and establish bounds on the winding number in relation to the order of non-locality. We additionally explore the physical challenges of realising non-local interfaces, where the extended connectivity of BNN interactions inherently introduces local perturbations at the interface, making the response sensitive to the precise boundary conditions. These perturbations can shift the frequencies of edge modes, remove their protection, or alter their number, undermining the predictions afforded by bulk winding numbers. This motivates a careful examination of the limitations of the SSH analogy in configurations where strong interface disorder is present, and of how the resulting interfacial sensitivity can be exploited.

\section{Non-Local Lattices}

\begin{figure}[h]
    \centering
    \includegraphics[width=1\linewidth]{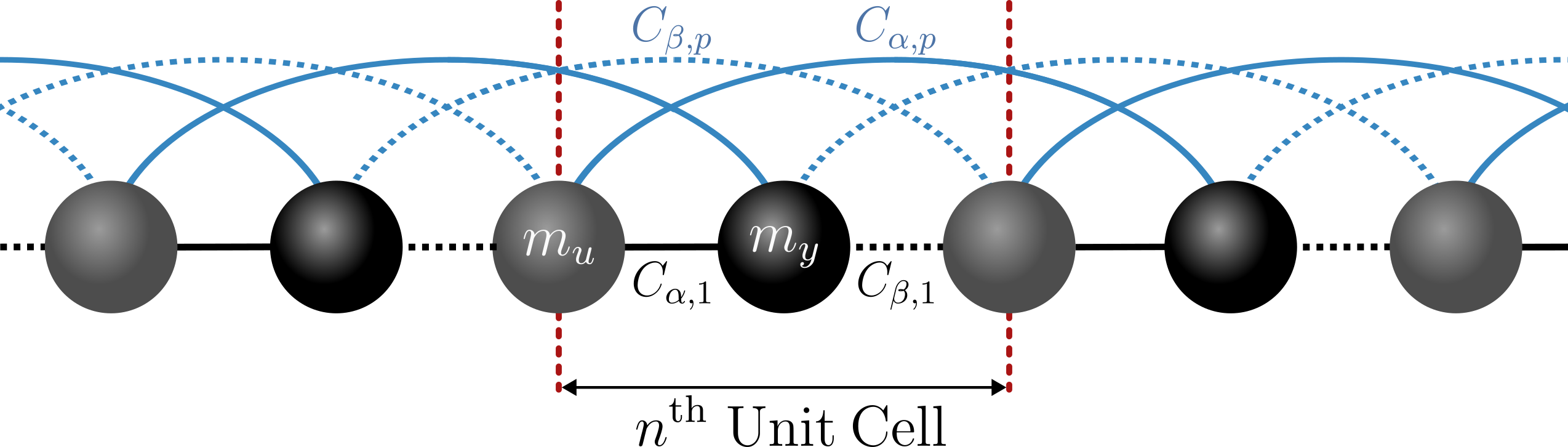}
    \caption{Non-local diatomic lattice of identical masses $(m=1)$ coupled by alternating spring stiffnesses, $C_\alpha$ and $C_\beta$. Nearest-neighbour (NN) connections are shown in black and whilst the BNN connections $(p=3)$ are blue, with solid and dashed lines distinguishing spring stiffnesses. The $n^\mathrm{th}$ unit cell is highlighted, with $u$ and $y$ distinguishing the two local environments.}
    \label{Fig: Dissimilar Sub-Lattice BNN}
\end{figure}

The canonical SSH mass–spring framework consists of a diatomic unit-cell where alternating strong and weak spring stiffnesses, between two identical masses, dictate the global topology via chiral symmetry protection (see appendix \ref{app: Chirality}). BNNs constitute a natural extension, given the design capabilities they exert over the system's dispersion. However, owing to the diatomic basis of the SSH model, BNN connections fall into two classes: even-order interactions that couple sites within the same sub-lattice and odd-order interactions that couple sites on different sub-lattices. These classes produce distinct topological responses. Even-order connections break the chiral symmetry that protects the SSH topology, limiting any topological distinction, so we restrict our attention to odd-order non-local connections. For further details see appendix \ref{app: Chirality}.

Figure \ref{Fig: Dissimilar Sub-Lattice BNN} depicts an example lattice with both NN and odd-ordered BNN connections, where the governing equations of motion are given by,
\begin{align}
    \begin{split}
        m\ddot{U}_n=&C_{\alpha,1} (Y_n-U_n)+C_{\beta,1}(Y_{n-1}-U_n)\\
        &+C_{\alpha,p}\left(Y_{n+\frac{p-1}{2}}-U_n\right)+C_{\beta,p}\left(Y_{n-\frac{p+1}{2}}-U_n\right),\\
        m\ddot{Y}_n =& C_{\beta,1}(U_{n+1}-Y_n)+C_{\alpha,1}(U_n-Y_n)\\
        &+C_{\beta,p}\left(U_{n+\frac{p+1}{2}}-Y_n\right)+C_{\alpha,p}\left(U_{n-\frac{p-1}{2}}-Y_n\right).
        \label{Eq: Dissimilar BNN EoM}
    \end{split}
\end{align}
The spring stiffnesses are denoted by $(C_{\alpha,p})$ or $(C_{\beta,p})$, where the subscript $(p)$ specifies the order of the connection. Consistent with the SSH framework, alternating strong and weak NN ($p=1$) connections are always included, while the spatial extent of the BNN interactions ($p>1$) can assume any value satisfying $p \in 2\mathbb{Z}+1$. The variables $U_n$ and $Y_n$ denote the mass displacements within the $n^\mathrm{th}$ unit cell, each of which is assumed to be time-harmonic, varying as $e^{\rm{i}\omega t}$, with unit mass, $m$. The equations of motion have been written to exclusively mediate interactions between NN and a single order of BNN. While the analysis is intentionally restricted to this case, the framework admits a generalisation to the inclusion of additional BNN connections \cite{EdgeMass2025}, provided lattice symmetries are preserved.

Utilising the Bloch ansatz, $e^{ikx}$, where $k$ is the wavevector, \eqref{Eq: Dissimilar BNN EoM} is recast as a dynamic matrix, returning a frequency domain generalised Eigenvalue problem in terms of $\omega^2$, 
\begin{widetext}
\begin{equation}
    \begin{bmatrix}
        \left(C_{\alpha,1}+C_{\beta,1}+C_{\alpha,p}+C_{\beta,p}\right)-m\omega^2 &
        -\left(C_{\alpha,1}+C_{\beta,1} e^{-2ik} + C_{\alpha,p} e^{ik(p-1)} + C_{\beta,p} e^{-ik(p+1)}\right) \\
        -\left(C_{\alpha,1}+C_{\beta,1} e^{2ik} + C_{\alpha,p} e^{-ik(p-1)} + C_{\beta,p} e^{ik(p+1)}\right) &
        \left(C_{\alpha,1}+C_{\beta,1}+C_{\alpha,p}+C_{\beta,p}\right)-m\omega^2
    \end{bmatrix}
    \begin{bmatrix}
        U_n \\
         Y_n 
    \end{bmatrix}=0.
\label{Eq: Dissimilar BNN Dynamic Matrix}
\end{equation}
\end{widetext}
To facilitate the analysis, we express the problem in terms of a reduced set of parameters while retaining the essential features of the system,
\begin{equation}
    \begin{bmatrix}
        \eta-m\omega^2 &
        \gamma \\
        \gamma^* &
        \eta-m\omega^2
    \end{bmatrix}
    \boldsymbol{U}
    =0.
\label{Eq: Reduced BNN Dynamic Matrix}
\end{equation}
The off-diagonal contribution, $\left[ \Re(\gamma) \boldsymbol{\sigma}_x-\Im(\gamma)\boldsymbol{\sigma}_y\right]$, where $\bm{\sigma}_x$and $\bm{\sigma}_y$ are the first two Pauli matrices, is equivalent to the Bloch Hamiltonian traditionally used in infinitely periodic SSH analysis \cite{Coutant2021, Chen2020}. However, this Hamiltonian relies on translational invariance and so cannot fully describe finite or disordered systems \cite{Zhang2023, Fukui2020} where extended BNN interactions modify the boundary conditions. We therefore adopt \eqref{Eq: Reduced BNN Dynamic Matrix}, which gives a complete and self-consistent description of the wave dynamics. Its eigenvalues follow directly from enforcing the solvability condition on the coefficient matrix returning,
\begin{equation}
    m\omega^2 = \eta \pm |\gamma|.
    \label{Eq: Odd BNN Eigenvals}
\end{equation}
In gapped systems, topological phenomena and localised states are supported by a band gap in the bulk dispersion, which creates a frequency range in which propagating bulk modes are forbidden. In the standard mass–spring system, a band gap opens when $C_\alpha \neq C_\beta$. However, with the introduction of BNN interactions the existence of the bulk band gap is now entirely governed by 
\begin{equation}
    |\gamma|=\left|C_{\alpha,1} + C_{\beta,1} e^{-2ik} + C_{\alpha,p} e^{ik(p-1)} + C_{\beta,p} e^{-ik(p+1)}\right|,
    \label{Eq: Complex trajectory}
\end{equation}
where the condition $\gamma = 0$ determines a closure of the band gap and consequently marks the onset of topological phase transitions. This competition between connection orders now permits band gaps to close away from high-symmetry points through bulk band crossings and Dirac-type points. The Zak phase is therefore ill-suited to characterising the global topology, a limitation compounded by its binary output \cite{Perez2018}.

\begin{figure*}[!t]
    \centering
    \includegraphics[width=0.8\textwidth]{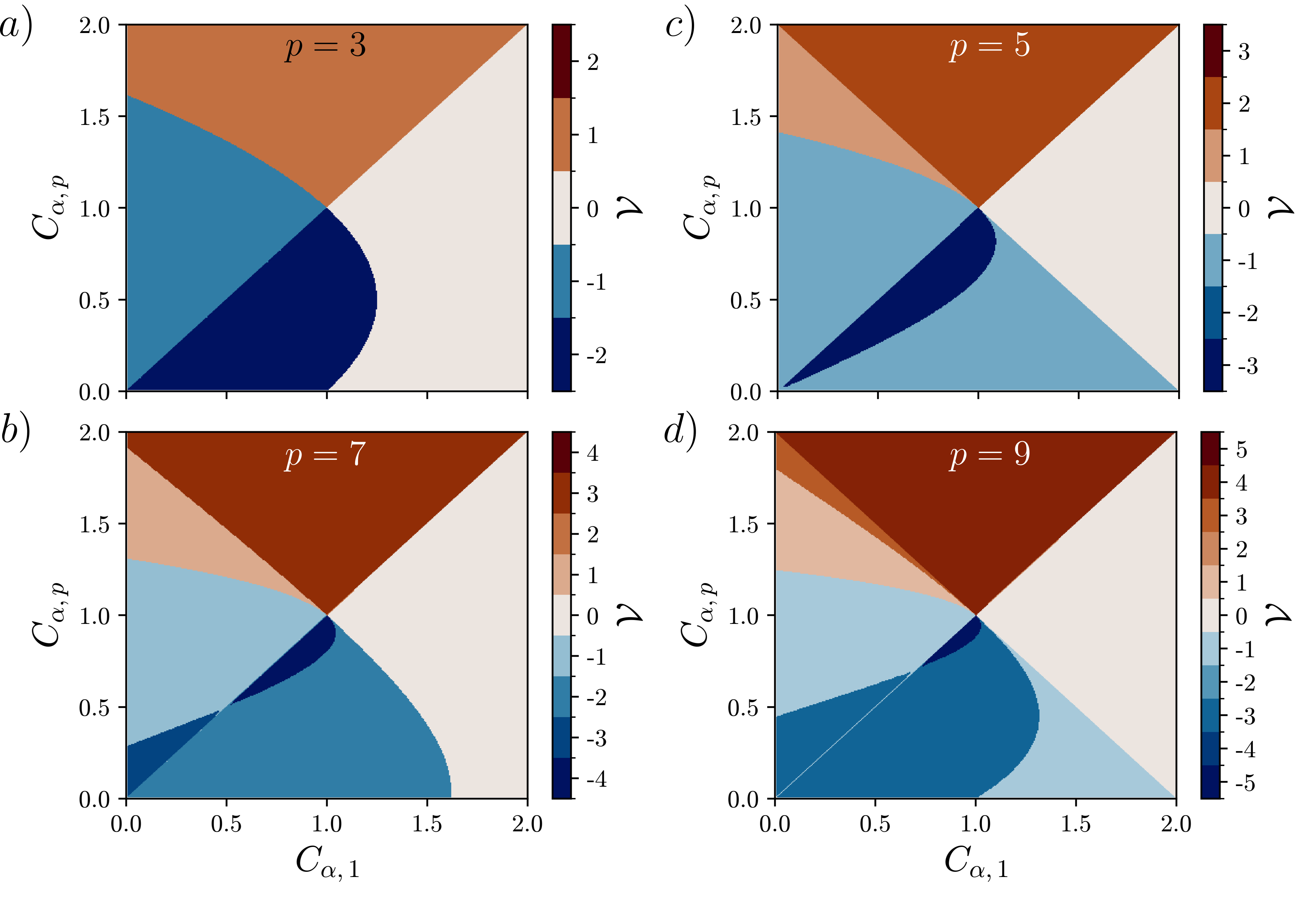}
    \caption{Winding number profiles for lattices with increasing orders of BNN connections. Here, $C_{\beta,1}=C_{\beta,p}=1$, whilst the $C_{\alpha}$ stiffness parameters are swept over the range $0 < C_{\alpha} \leq 2$. The plot colours denote the corresponding integer-quantised winding numbers, with the sign indicating whether the complex trajectory $\gamma$ encircles the origin clockwise $(\mathcal{V}<0)$ or anti-clockwise $(\mathcal{V}>0)$. Boundaries between regions of differing winding correspond to closures of the bulk band gap, marking the onset of topological phase transitions. The range of accessible winding numbers increases with the spatial extent of the BNN connections, specified by the value of $p$ indicated at the top of each panel.}
    \label{Fig: Dissimilar BNN Winding}
\end{figure*}

The winding number provides a more complete classification of non-local lattice topology \cite{Chen2020}, counting how many times the complex coupling trajectory, $(\gamma)$ encircles the origin as the Brillouin zone is traversed,
\begin{align}
\begin{split}
    \mathcal{V} = \frac{1}{2\pi} \oint_{\mathrm{BZ}} \frac{\partial}{\partial k} \left( \Im \ln \gamma(k) \right) dk
    = \frac{1}{2\pi i} \oint_{\mathrm{BZ}} \frac{\gamma'(k)}{\gamma(k)} dk.
    \label{Eq: Winding Number}
\end{split}
\end{align}

Figure \ref{Fig: Dissimilar BNN Winding} presents the winding number profiles for lattices with increasing orders of BNN connections. In each plot the stiffness parameter space is swept, keeping $C_{\beta,1}=C_{\beta,p}=1$, while the $C_\alpha$ spring stiffnesses are varied. The resulting winding numbers encode both the number and direction of encirclements of the origin by the complex trajectories, with the sign distinguishing clockwise ($\mathcal{V}<0$) windings from anticlockwise ($\mathcal{V}>0$). Every boundary between winding numbers marks a topological phase transition. 

The Winding number exposes the richer topology of non-local lattices that is not accessible in local systems. The added complexity of BNN interactions modifies the analytic structure of the dynamical matrix across the Brillouin zone, introducing higher-index exponential terms in the off-diagonal complex coupling $\gamma$. This increased polynomial index consequently facilitates higher-order windings beyond those present in NN models. For each lattice structure, the spatial extent of its couplings imposes a bound on the attainable winding numbers, limiting the maximum topological order that can be supported. We establish this bound using contour integration residue calculus, drawing on the argument principle, which relates the winding number of a closed contour around the origin to the difference between the numbers of enclosed zeros and poles of the associated complex function, counted with multiplicity \cite{Needham2023}. 

Adopting the standard anti-clockwise convention, $\gamma$ can be re-expressed in terms of the Bloch-invariant mapping $z = e^{i k \ell}$ of unit radius $|z| = 1$, where $\ell=2$ denotes the lattice spacing, returning an ordinary complex polynomial ($\Gamma(z)$),
\begin{align}
\begin{split}
    \Gamma(z)&=z^{\frac{(p+1)}{2}}\gamma(z) \\ &= C_{\alpha,1}z^{\frac{(p+1)}{2}} + C_{\beta,1}z^{\frac{(p-1)}{2}} + C_{\alpha,p} z^{p} + C_{\beta,p}.
    \label{Eq: Ordinary Complex trajectory (mapped)}    
\end{split}
\end{align}
Substitution of $\Gamma(z)$ into \eqref{Eq: Winding Number}, returns the following expression for the winding of $\gamma(z)$. Analogously, this is the argument principle, 
\begin{align}
\begin{split}
    \mathcal{V} &= \frac{1}{2\pi i} \oint_{\mathcal{C}} \frac{\gamma(z)'}{\gamma(z)}\\[9pt]
    &= \underbrace{\frac{1}{2\pi i} \oint_{\mathcal{C}} \frac{\left(z^{\frac{(p+1)}{2}}\gamma(z)\right)'}{z^{\frac{(p+1)}{2}}\gamma(z)} dz}_{\mathrm{Zeros}}
    -\underbrace{\frac{1}{2\pi i} \oint_{\mathcal{C}}\frac{(p+1)}{2}\frac{1}{z} dz}_{\mathrm{Poles}}.
    \label{Eq: Argument Principle}
\end{split}
\end{align}
where we have highlighted the expressions for the zeros and poles. For further details see appendix \ref{app: winding bounds}. 

There exists a single pole at $z=0$ with residue $\frac{p+1}{2}$. Similarly, the number of zeros are determined by the degree $z^{\frac{p+1}{2}}\gamma(z)$, which spans powers from $z^{0}$ to $z^{p}$. Counting multiplicity, up to $p$ zeros can be admitted, with the exact number depending on the specific lattice coefficients, potentially reducing under degeneracies to fewer zeros including the limiting case of none. Therefore the span of admissible winding numbers is governed solely by the maximum order of spatial connections within the lattice structure, ranging between
\begin{equation}
    \mathcal{V}_{\max} = p - \frac{p+1}{2} = \frac{p-1}{2}, \qquad
    \mathcal{V}_{\min} = 0 - \frac{p+1}{2} = -\frac{p+1}{2}.
    \label{Eq: Winding Bounds}
\end{equation}
This provides a theoretical basis for designing SSH-inspired lattice structures capable of supporting multiple topologically protected boundary states.

Within the attainable bounds, the realised winding number is determined by the relative strengths of the spring stiffnesses. Incorporating non-local connections enlarges the parameter space, with each order of spatial connection contributing its own spring stiffnesses. This affords greater tunability of both the lattice dynamics and its topology. Small variations in these parameters drive transitions between neighbouring phases, subject to the bound imposed by the lattice connectivity (see Figure~\ref{Fig: Dissimilar BNN Winding}). Rouché’s theorem offers a means of examining this variation \cite{Needham2023}, showing how the winding number evolves when a single coupling order dominates. In the limit where the local, NN stiffnesses dominate over all higher-order connections, the topology of the lattice reduces to that of the canonical SSH model, returning winding numbers of $|\mathcal{V}|=\{0,1\}$. Details are given in Appendix \ref{app: winding bounds}. However, despite the admissible range of winding numbers, not all values within this interval are necessarily accessible owing to the restricted orders of spatial connection. With only NN and a single order of BNN connections present, the resulting complex polynomial contains a sparse distribution of powers, particularly as the range of BNN interactions increases. Consequently, certain intermediate winding states cannot be realised. This is evident in Figure~\ref{Fig: Dissimilar BNN Winding}d), where the state $|\mathcal{V}|=2$ is absent despite lying within the admissible winding range.


\section{Non-Local Interfaces}

The dual control over topological and dynamic wave phenomena afforded by non-local connections substantially broadens the theoretical scope for practical application, potentially facilitating novel solutions across a range of wave-based technologies, including sensing, wave-guiding, and imaging and signal enhancement \cite{Sun2024, Sun2025, Miniaci2021}. However, this potential is tempered by the complexity introduced by BNN connections, which makes the precise choice of boundary conditions and lattice termination critical to the challenge of designing non-local interfaces that preserve the topology. The precise configuration of discrete non-local interfaces presents a particular challenge. BNN couplings span multiple sites, raising questions about which bonds to retain, truncate or modify when joining two lattices, and how to tune the corresponding stiffnesses across the boundary to preserve the topology.

The bulk-boundary correspondence relates the behaviour at an interface to the bulk topology, with the absolute difference between bulk windings predicting the number of interface states formed at the boundary between two infinite half-spaces,
\begin{equation}
    \mathcal{N}=|\mathcal{V}_{\rm{L}}-\mathcal{V}_{\rm{R}}|,
    \label{Eq: Nunmber of Interface States}
\end{equation}
where the $\rm{L}$ and $\rm{R}$ subscripts denote the left and right half spaces, respectively. However, this prediction is an idealisation that assumes an abrupt transition between the two lattices. In the full dynamical mass-spring framework, the eigenfrequencies of the localised interface states remain highly sensitive to how the boundary is physically realised. For NN systems this distinction is relatively minor, as the interface is confined to a single disrupted coupling between adjacent masses, whereas in non-local lattices multiple coupling orders are involved at the boundary.

Multiple coupling orders replace the sharp discontinuity of a conventional interface with a finite transition region, whose extent is determined by the largest connection order. This region is not uniquely defined, as the retained coupling orders and their assigned stiffnesses each produce a distinct local environment, breaking local symmetries and introducing perturbations that deviate from the idealised bulk description. These effects become more prevalent in finite lattices, where, as $p$ approaches the domain size, the distinction between bulk and boundary becomes increasingly blurred, disrupting bulk periodicity and complicating the application of bulk-boundary correspondence. Despite this, dislocated and disordered interfaces have been shown to support robust topological modes under perturbations \cite{ammari2020, ammari2022, Craster2023}, while also highlighting the dependence of mode localisation and eigenfrequencies on the nature and extent of the disorder. Thus, the interface itself can become a design choice rather than a fixed boundary condition, raising the question of how best to configure a non-local interface.

\begin{figure}[h!]
    \centering
    \includegraphics[width=0.8\linewidth]{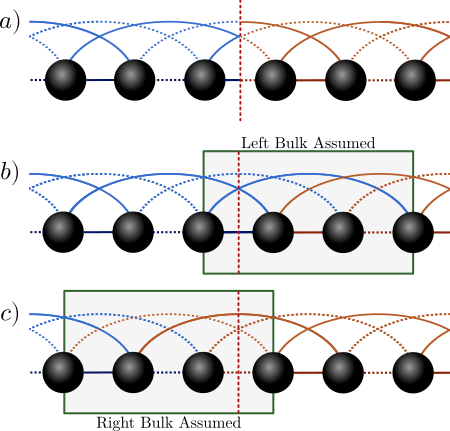}
    \caption{a) Non-Local SSH-analogue model, depicting the interface between two bulk domains. The left (blue) and right (red) lattices form a discontinuity at the boundary, where the interfacial springs simultaneously inherit stiffness terms associated with both bulk materials. This construction highlights the design problem of determining appropriate interfacial coupling values that consistently connect the two bulk domains while preserving the desired properties. (b-c) The interfacial coupling structure in a lattice with NN and $3\mathrm{rd}$ NN spatial connections, whereby the spring stiffnesses are inherited from the left (b) and right (c) bulk domains, respectively. The vertical dashed line denotes the defined interface, with the green box highlighting the finite extent of the discrete interface region due to the interfacial spring connections.}
    \label{Fig: Assumed Stiffness BNN Interface}
\end{figure}

We consider interface configurations analogous to the discrete SSH model, with the transition between two topologically distinct bulk regions imposed directly through the spring stiffnesses (see Fig.~\ref{Fig: Assumed Stiffness BNN Interface}a). In this SSH-analogue model, all interfacial springs are defined to exclusively adopt the lattice properties of the left or right bulk, respectively. This choice is intentional, as it conforms to the convention of the standard mass-spring SSH model while also minimising the width of the interface to a span of $p$ masses, rather than the $2p$ masses it could otherwise involve. A schematic of the adopted interfacial configurations are shown in Figures \ref{Fig: Assumed Stiffness BNN Interface}(b-c). In the canonical SSH model, an interface typically joins two chiral-symmetric lattices that share the same bulk dispersion and differ only in their topological phase. The non-local couplings considered here relax this restriction, permitting interfaces between dissimilar bulk lattices with different winding numbers and dispersion relations. This broadens the scope of topological interface design beyond conventional SSH configurations.

\begin{figure*}[!t]
    \centering
    \includegraphics[width=\textwidth]{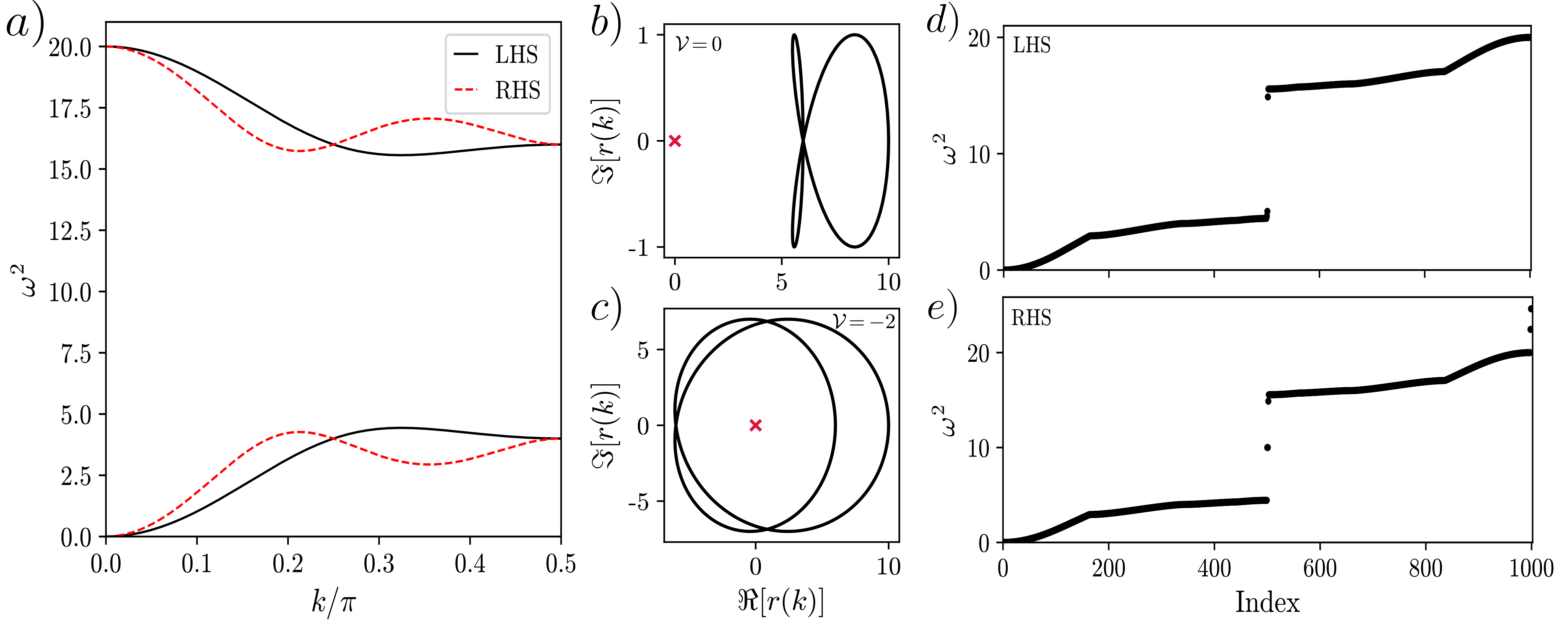}
    \caption{(a) Bulk dispersion profiles for the left (solid black) and right (dashed red) bulk lattices, with spring stiffnesses defined in Table \ref{Tab: Dissimilar Bulk Stiffnesses}. Both dispersion relations are designed to share a common mid-gap at $\omega^2=10$. The corresponding complex trajectories of the (b) left and (c) right bulk windings constituent to the interface. The red cross denotes the origin, while the calculated winding number for each trajectory is indicated by the text in the corner of each panel. Panels (d) and (e) show the raw, unordered eigenvalues for the interface study, with the interfacial spring stiffnesses taken exclusively from the left and right lattice, respectively. The finite domain spans $N=1000$ masses, with the interface initially defined between masses $N=499$ and $N=500$ prior to assigning the stiffnesses.}
    \label{Fig: Interface BNN Study}
\end{figure*}

We exploit this dual control over dynamic and topological phenomena by interfacing two bulk lattices that differ in both dispersion and topology, an approach with precedent in dissimilar NN waveguides \cite{Morini2025}. The stiffness parameters are chosen so that the left (L) and right (R) bulk materials share a common band gap centred on the same mid-gap frequency, which is enforced by $\eta_{\rm{L}}=\eta_{\rm{R}}$, with $\eta$ as defined in Eq.~\eqref{Eq: Reduced BNN Dynamic Matrix}. The left half-space is dominated by a NN connection order, admitting a winding number of $\mathcal{V}=0$, whilst the right half-space is dominated by a $p=3$ BNN connection, admitting a winding number of $\mathcal{V}=-2$. Figure \ref{Fig: Interface BNN Study}(a-c) depicts both the bulk dispersion profiles and the complex trajectories associated with the bulk windings. The corresponding stiffness values are detailed in Table~\ref{Tab: Dissimilar Bulk Stiffnesses}, with the stiffnesses labelled as in Figure \ref{Fig: Dissimilar Sub-Lattice BNN}.

\begin{table}[h!]
\centering
\renewcommand{\arraystretch}{1.4} 
\setlength{\tabcolsep}{14pt}      
\begin{tabular}{|c|c|c|c|}         
\hline
$p$ (Connection order) & $p=1$ & $p=3$ \\
\hline
$C_{\alpha, \text{L}}$  & 7   & 1   \\
$C_{\beta, \text{L}}$   & 1   & 1  \\
\hline
$C_{\alpha, \text{R}}$  & 1 & 1    \\
$C_{\beta, \text{R}}$   & 1 & 7   \\
\hline
\end{tabular}
\caption{Spring stiffnesses for the left-hand side (L) and right-hand side (R) lattice half-spaces with NN ($p=1$) and third-nearest-neighbour ($p=3$) couplings. $\alpha$ and $\beta$ define the two spring stiffness for each coupling order in the diatomic lattice basis, as depicted in Figure \ref{Fig: Dissimilar Sub-Lattice BNN}.}
\label{Tab: Dissimilar Bulk Stiffnesses}
\end{table}
A modal analysis of a finite lattice of $N=1000$ masses is performed to investigate localised interface states, with the interface (vertical red dashed line in Fig.~\ref{Fig: Assumed Stiffness BNN Interface}) defined on the spring connections between masses $N_\mathrm{Int}=\{499,500\}$. At the lattice ends, BNN connections are truncated, introducing a small degree of local disorder whose influence on the interface states is negligible given the connection orders and domain size. The raw unordered eigenvalues obtained for the left- and right-assumed interfacial configurations are presented in Fig.~\ref{Fig: Interface BNN Study}(d-e).

In both studies, the finite domain size gives rise to a left-hand localised edge state at $\omega^2=14.880$. Additionally, local disorder around the interface configuration yields defect states above the bulk bands, particularly for the right-bulk-assumed interface. Distinct from these, and in agreement with \eqref{Eq: Nunmber of Interface States}, two localised interface states are observed within the common band gap. Their behaviours, however, differ markedly, yielding distinct eigenfrequencies and mode shapes. 

\begin{figure*}
    \centering
    \includegraphics[width=0.9\linewidth]{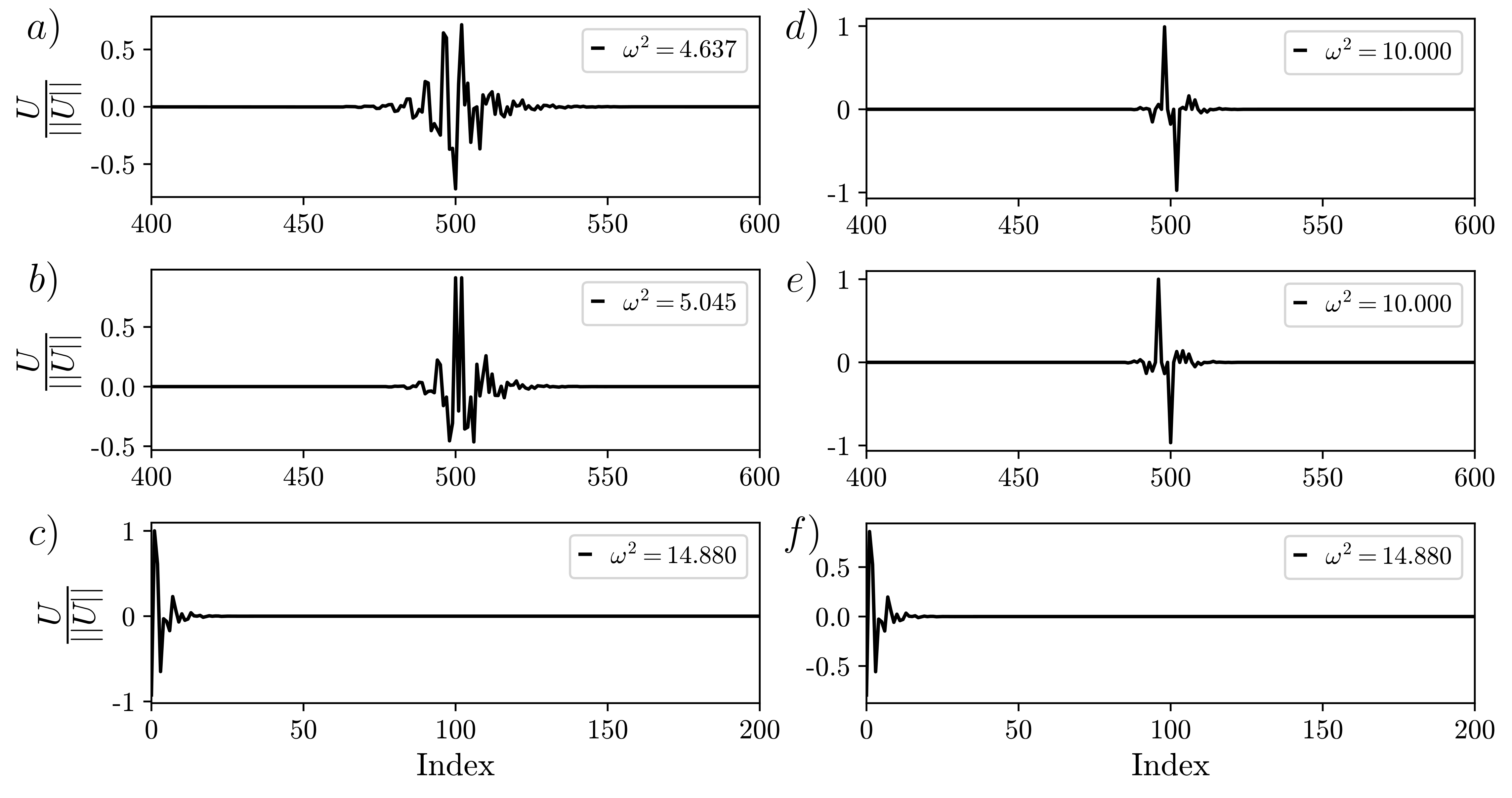}
    \caption{Eigenmodes of the localised states identified within the finite band gaps in Figure \ref{Fig: Interface BNN Study}. Each eigenmode is normalised by the maximum absolute amplitude across all system eigenmodes (note different $y$-axis scales), with the resulting amplitude providing a measure of the localisation strength. Panels (a–c) show the eigenmodes obtained when the interfacial spring stiffnesses are assigned from the left-hand bulk lattice, revealing two interface states and one edge state. Conversely, panels (d–f) show the eigenmodes for interfacial stiffnesses assigned from the right-hand bulk lattice, exhibiting two mid gap localised interface states and a single edge state. In the top right of each panel is the frequency of the plotted eigenmode.}
    \label{Fig: Dissimilar Interface Bulk Eigenmodes}
\end{figure*}

When the interfacial configuration is adopted from the right bulk, which pertains to a larger winding amplitude, a degenerate pair of mid-gap states are observed. The eigenmodes are strongly localised about the interface (see Figure \ref{Fig: Dissimilar Interface Bulk Eigenmodes}), where the asymmetry in the eigenmode profile is a by-product of the exact interfacial configuration. Mathematically, this introduces inconsistencies in the on-diagonal stiffness contributions resulting in local chiral symmetry breaking. Conversely, when the interfacial stiffnesses are adopted from the left bulk, the localised states reside closer to the lower bulk band. The corresponding normalised eigenmodes exhibit a reduced amplitude and a weaker overall localisation, which can be attributed to the displacement of these states from the mid-gap. Analytical approaches such as the transfer matrix method and root-finding techniques prove challenging within this non-local formalism, since the dimensionality of the problems grows with the range of the BNN. Alternative numerical methods, such as high-frequency homogenisation or the Thouless formula for the Lyapunov exponent, can instead be used to characterise the localisation strength at frequencies across the band gap \cite{Hu2025, Li2024, Craster2010}.

The disparity between the localised modes, despite their common topological origin, emphasises the sensitivity of the interface states to the exact boundary configuration imposed when constructing a non-local interface. In discrete non-local systems, the extent to which topological predictions translate into localised states is limited by the disorder that the interface configuration introduces. The interface can be configured to favour states closer to mid-gap, enhancing their robustness to disorder. However, bulk topological predictions alone do not necessarily guarantee the existence or localisation of interface states when the interfacial region is disordered, with the specific interface configuration potentially yielding a number of emergent interface states that deviates from the idealised prediction.

\begin{figure*}[!t]
    \centering
    \includegraphics[width=0.8\textwidth]{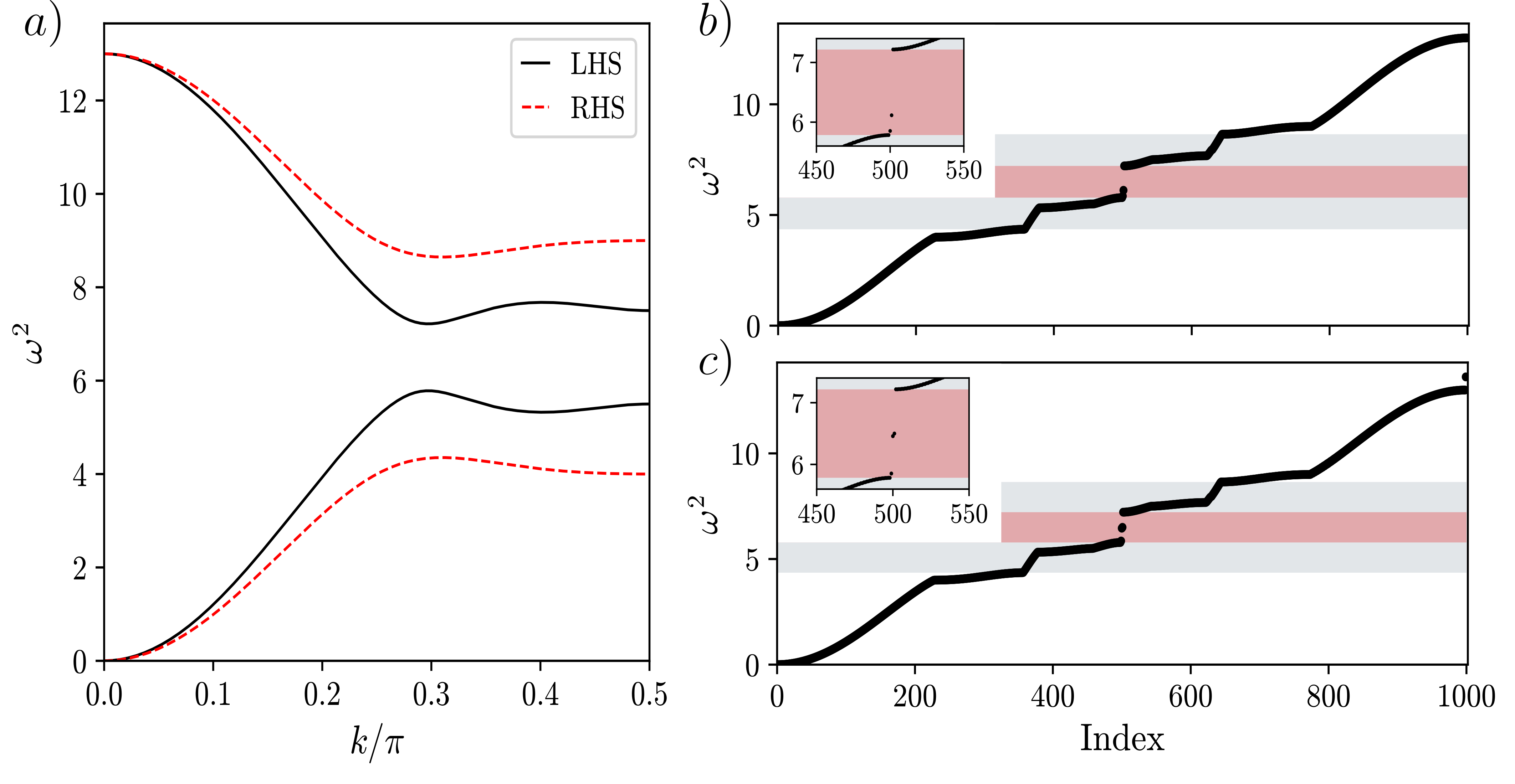}
    \caption{(a) Dispersion curves of the left and right bulk lattices, designed to share a common mid gap at $\omega^2 = 6.5$. The left bulk exhibits a narrower band gap spanning $5.784 < \omega^2 < 7.216$, whereas the wider band gap of the right bulk spans $4.353 < \omega^2 < 8.647$. Raw, unordered eigenvalues for the interface between two dissimilar bulk lattices, with interfacial spring stiffnesses assigned exclusively from the right (b) and left (c) bulk lattices. The finite domain comprises $N=1000$ masses, with the interface initially defined between masses $N=499$ and $N=500$, prior to assigning the interfacial stiffnesses. The red and grey shaded regions denote the band gaps of the left and right bulk lattices, respectively. Insets (upper left) enlarge the eigenvalues within the common band gap.}
    \label{Fig: Dissimilar Bulk Winding Breakdown}
\end{figure*}

A specific example of the bulk winding number failing to predict the number of boundary states occurs at the interface between the two bulk domains with stiffness values as given in Table \ref{Tab: Dissimilar Bulk Winding Breakdown}.
\begin{table}[h!]
\centering
\renewcommand{\arraystretch}{1.5} 
\setlength{\tabcolsep}{14pt}      
\begin{tabular}{|c|c|c|c|}         
\hline
$p$ (Connection order) & $p=1$ & $p=3$ \\
\hline
$C_{\alpha, \text{LHS}}$  & 1.75   & 0.75   \\
$C_{\beta, \text{LHS}}$   & 2   & 2  \\
\hline
$C_{\alpha, \text{RHS}}$  & 3.5 & 1    \\
$C_{\beta, \text{RHS}}$   & 1 & 1   \\
\hline
\end{tabular}
\caption{Spring stiffnesses for the left (LHS) and right-hand side (RHS) dissimilar bulk lattices with NN ($p=1$) and third-nearest-neighbour ($p=3$) couplings. $\alpha$ and $\beta$ define the two spring stiffness for each coupling order in the diatomic lattice basis, as depicted in Figure \ref{Fig: Dissimilar Sub-Lattice BNN}.}
\label{Tab: Dissimilar Bulk Winding Breakdown}
\end{table}
The stiffness parameters are deliberately chosen such that the right-hand bulk exhibits a wider band gap that entirely contains the band gap of the left-hand bulk, yielding a common band gap spanning $5.784<\omega^2<7.216$, as depicted in Figure \ref{Fig: Dissimilar Bulk Winding Breakdown}a). This choice of stiffnesses also produces distinct bulk winding numbers of $\mathcal{V}_{\mathrm{L}}=-2$ and $\mathcal{V}_{\mathrm{R}}=0$, from which $\mathcal{N}=2$ interface states are predicted within the common finite band gap.

\begin{figure*}[!t]
    \centering
    \includegraphics[width=0.9\textwidth]{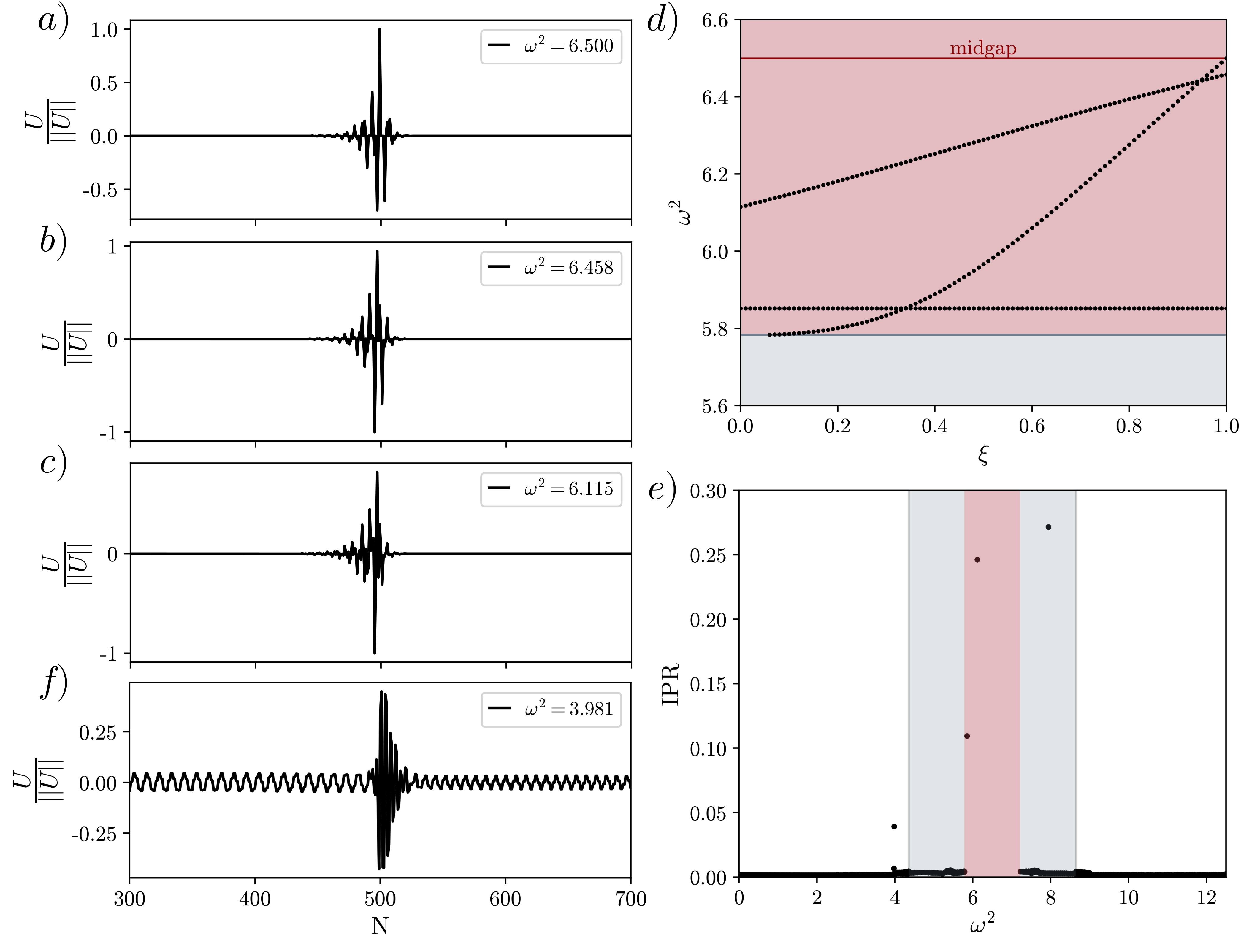}
    \caption{(a-c) Normalised eigenmodes of the localised interface states within the common band gap, with interfacial spring stiffnesses assigned from the left (a,b) and right (c) bulk lattices. The eigenvalue of each mode is given in the upper-right corner of its panel. (d) Tracked localised eigenmodes (dots) within the band gap as $0 \leq \xi \leq 1$ is varied, mapping the transition between the interface configurations discussed in Fig.~\ref{Fig: Dissimilar Bulk Winding Breakdown}. The horizontal red line marks the mid-gap, as the $y$-axis range is restricted. (e) Inverse participation ratio (IPR) corresponding to Fig.~\ref{Fig: Dissimilar Bulk Winding Breakdown}b), with the horizontal red dashed line marking the IPR tolerance of five times the average IPR. In (d,e), the red and grey shaded regions indicate the common band gap (the left bulk band gap) and the broader band gap of the right bulk, respectively. (f) Leaky resonance localised around the interface, with propagating wavefronts of different amplitudes in the two dissimilar half-spaces away from the interface. All eigenmodes are normalised by the maximum absolute amplitude across all system eigenmodes.}
    \label{Fig: IPR and Eigenmodes}
\end{figure*}

The raw, unordered eigenvalues (Fig.~\ref{Fig: Dissimilar Bulk Winding Breakdown}(b-c)) show that both interfacial configurations support an edge state at $\omega^2=5.852$. Beyond this, the two configurations differ. Adopting the interfacial stiffnesses from the right bulk yields a single localised interface state at $\omega^2=6.115$ (Fig.~\ref{Fig: Dissimilar Bulk Winding Breakdown}(b)), whereas adopting those from the left bulk yields two interface states within the common band gap (Fig.~\ref{Fig: Dissimilar Bulk Winding Breakdown}(c)). The corresponding interfacial eigenmodes are presented in Figures \ref{Fig: IPR and Eigenmodes}(a-c), detailing skewed localisation profiles governed primarily by the distinct bulk band-gap structures on either side of the interface. 

To capture the influence of the interfacial configuration on the emergence of interface states, a continuous mapping is performed between the two configurations considered, weighting the interfacial stiffness contributions from each bulk medium. A scaling parameter $0\leq\xi\leq1$ is introduced,
\begin{equation}
C_{\rm{int}} = \xi C_{\rm{L}}+(1-\xi)C_{\rm{R}}.
\end{equation}
The limiting cases of $\xi=0$ and $\xi=1$ recover the previously discussed interface structures, with the interfacial stiffnesses inherited from the right and left bulk, respectively. As $\xi$ increases from zero to unity, the edge state remains fixed in frequency (Figure \ref{Fig: IPR and Eigenmodes}d)). Conversely, the interface-state frequencies vary continuously with the interfacial configuration, displaying both linear and non-linear trends. At $\xi=0$, a single interface state lies within the common band gap, while the second has merged into the lower bulk dispersion band. With increasing $\xi$, the second state detaches from the lower bulk band and localises within the gap, evolving along a higher-order non-linear trajectory. Under this deterministic perturbation of the interface, the delocalisation of the interface states can be observed directly. The modes track and hybridise into the bulk bands, visualising the breakdown of the bulk-winding prediction while also highlighting the sensitivity to the interfacial configuration.

As localised modes approach the bulk bands, hybridisation between the localised and extended bulk states increases, with its extent influenced by several factors including the bulk dispersion, the spring stiffnesses, and the order of the BNN connections. In the raw eigenvalue spectrum, only localised states within the common band gap can be identified. We therefore use the inverse participation ratio (IPR) to quantify the localisation of each eigenmode. As the IPR scales with the domain size, its absolute value has no universal meaning, so the output is treated as a guide for identifying candidate states, and the explicit eigenmodes of interest are then examined on a case-by-case basis. The IPR is defined as
\begin{equation}
\mathrm{IPR} = \frac{\sum_n \left|\boldsymbol{u}_n\right|^4}{\left(\sum_n \left|\boldsymbol{u}_n\right|^2\right)^2},
\label{Eq: IPR}
\end{equation}
where $\boldsymbol{u}_n$ is the displacement of the $n^{\rm{th}}$ mass in a given eigenmode, irrespective of its sublattice. Figure \ref{Fig: IPR and Eigenmodes}(e) presents the IPR values corresponding to Figure \ref{Fig: Dissimilar Bulk Winding Breakdown}(b), highlighting the previously identified edge and interface states within the common band gap. The largest IPR belongs to an additional edge state that the raw eigenvalue picture did not reveal, sitting solely within the right bulk’s band gap (grey region).

Notably, a fourth eigenmode has an IPR above that of the extended bulk states, occurring within the lower bulk bands at $\omega^2 = 3.981$. Unlike the previously identified edge and interface states, this mode lies within the bulk spectrum. While this eigenmode does not represent a conventional band-gap localised state, its enhanced localisation relative to the bulk modes indicates a distinct form of localisation that differs from the typical gap-supported states, as shown in Figure \ref{Fig: IPR and Eigenmodes}f). Although the mode is concentrated around the interface at $N\approx500$, it does not exhibit the exponential decay characteristic of the previously identified interface states. Instead, oscillatory wave propagation is observed on either side of the interface, with distinct propagation characteristics in the two bulk regions. This shows the mode is only partially confined, coupling to propagating bulk states while remaining enhanced at the interface. Such behaviour is characteristic of a leaky resonance rather than a true band gap interface state, and can be investigated further with a local density of states analysis \cite{Khattou2022}.

\section{Conclusions}
Non-local interfaces pose a design challenge because the explicit interfacial configuration inevitably introduces local perturbations. The resulting interface states are therefore sensitive to both the bulk topology and the configuration adopted at the boundary between dissimilar non-local media. While the bulk-boundary correspondence predicts the number of interface states, it does not capture their frequencies or localisation characteristics, which depend on the specific interfacial configuration. In particular, variations in the interface can induce hybridisation with the bulk bands, altering the character of the resulting modes. The finite spatial extent of the BNN connections in a discrete lattice further complicates a robust topological interpretation, such that bulk topology should be treated as a guide to the existence of interface states rather than a definitive description of their behaviour. Nevertheless, the interfacial configuration provides an additional degree of freedom for controlling the resulting modes, allowing their spectral position to be tuned within the common band gap, with a corresponding influence on their spatial localisation.

Extending a discrete lattice to include BNN couplings enriches its physics, enabling greater control over both dynamic and topological behaviour. It facilitates a broader range of dispersion profiles, in which both the bulk dispersion and the band-gap structure can be predictably engineered. At the global level, BNN connections lead to a diverse topological landscape, supporting a wider range of winding numbers that is bounded by the lattice connectivity but sensitive to the precise spring stiffnesses. Together, this dual control of dynamic and topological phenomena allows emergent localised states to be tailored not only in number, but also in character. We have shown that the interfacial configuration controls where the localised states sit within the band gap, while the boundary between media with dissimilar bulk dispersion can be designed to skew their localisation profiles, giving rise to asymmetric wave phenomena. We envisage that these findings will provide a practical basis for interface design in BNN lattices, where the choice of bulk properties and interfacial configuration can be used to tailor the number, spectral position and spatial character of localised states.

\begin{acknowledgements} 
\noindent RGE acknowledges the financial support of Defence Science and Technology Laboratory (Dstl). GJC acknowledges financial support from the EPSRC via the META4D Programme Grant (No. EP/Y015673/1). All data created during this research is available upon reasonable request to the corresponding author. ‘For the purpose of open access, the author has applied a ‘Creative Commons Attribution (CC BY) licence to any Author Accepted Manuscript version arising from this submission’.
\end{acknowledgements}

\appendix

\section{Chirality and Quantisation \label{app: Chirality}}

Topology is often characterised using invariants of various forms, such as Chern classes for complex vector bundles over smooth manifolds, or winding numbers and Berry-phase-based invariants in periodically driven or crystalline systems. Topological invariants provide a global classification of physical systems that are insensitive to local perturbations, quantifying properties that remain unchanged under continuous deformations, provided the system’s symmetry constraints and spectral gap are preserved. These invariants are typically integer-quantised, with different integer values corresponding to distinct topological phases of matter that cannot be transitioned between without undergoing a phase transition.

\begin{figure}[h]
    \centering
    \includegraphics[width=0.7\linewidth]{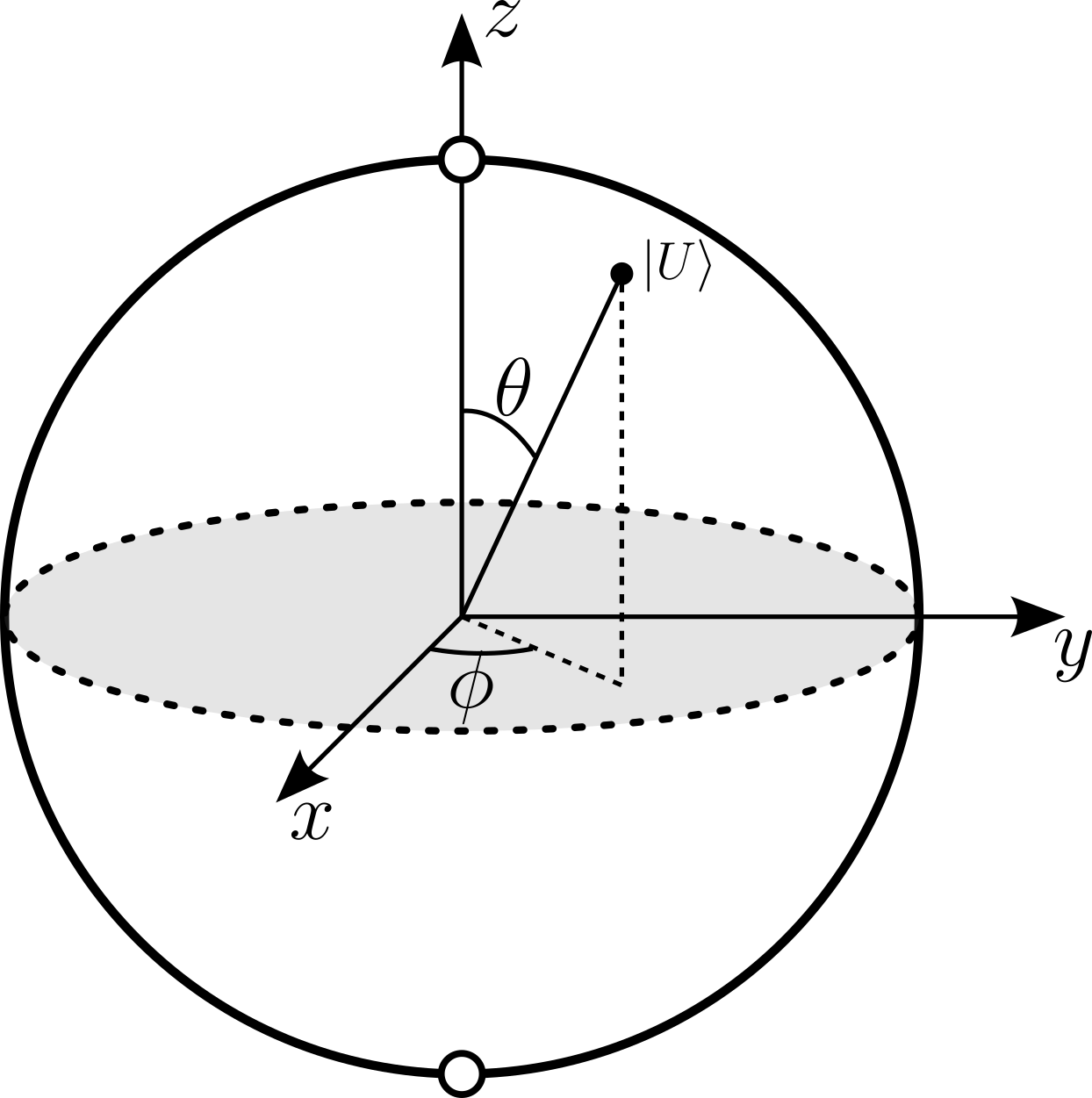}
    \caption{Schematic of the Bloch sphere, highlighting both the azimuthal $(\phi)$ and polar angle $(\theta)$ for Eigenmode $|U\rangle$ and the coordinate axis $(x,y,z)$. The chirality defined plane with polar angle $\pi/2$, is depicted by the grey circle.}
    \label{Fig:Bloch Sphere}
\end{figure}

Consider a general two-level Hamiltonian for a periodic medium, $\hat{H}(\mathbf{k})=\textbf{h}(\mathbf{k})\cdot \boldsymbol{\sigma}=h_x\sigma_x + h_y\sigma_y+h_z\sigma_z$, where $\textbf{h}(\mathbf{k})\in \mathbb{R}^3$. By Bloch’s theorem, its eigenstates take the form $|\psi_{\mathbf{k}}\rangle = e^{i\mathbf{k}\cdot \mathbf{r}} |u_{\mathbf{k}}\rangle$ where $|u_{\mathbf{k}}\rangle$ is lattice periodic, and the Hamiltonian satisfies the periodic gauge condition $\hat{H}(\mathbf{k}+\mathbf{G})=\hat{H}(\mathbf{k})$, where $\mathbf{G}$ is a reciprocal lattice vector. Accordingly, topological invariants may be evaluated over closed contours across the Brillouin zone without loss of generality.

With no symmetry constraints imposed, the Bloch eigenvectors,  
\begin{equation}
    |u(k)\rangle = 
    \begin{bmatrix}
    \cos\left(\frac{\theta}{2}\right) \\
     e^{i\phi}\sin\left(\frac{\theta}{2}\right)
    \end{bmatrix},
    \label{Eq: Bloch Sphere Vectors}
\end{equation}
span the full Bloch sphere of the unrestricted parameter space of $\textbf{h}$, exploring both the polar angle $\theta=\arccos\left(h_z/|\textbf{h}|\right)$ and the azimuthal angle $\phi=\arctan\left(h_y/h_x\right)$ (Figure \ref{Fig:Bloch Sphere}). Any quantity accumulated around the Brillouin zone is therefore unrestricted and varies continuously under smooth deformations of the Hamiltonian.

Imposing a chiral symmetry condition, the Hamiltonian is required to satisfy the anti-commutation relation $(\Gamma \hat{H} \Gamma^{-1} = -\hat{H})$, where for a two-level system $\Gamma = \sigma_z$. The chirality operator $(\Gamma)$ must be both unitary and hermitian, $\Gamma^{\dagger}\Gamma = \Gamma^2=\mathbb{I}$. Enforcing this condition constrains the Hamiltonian to $h_z = 0 \:\forall \: k$, fixing the polar angle to $\theta = \pi/2$. Consequently, as the closed-loop contour is traversed, namely the Brillouin zone in a periodic medium, the eigenstates are restricted to a well-defined trajectory on the equatorial plane of the Bloch sphere.

As the contour forms a closed loop in parameter space, the eigenstate must return to itself up to a phase equivalence consistent with the periodic gauge condition. In general, invariants such as the Berry phase,
\begin{equation}
\gamma = \oint \mathbf{A}(\mathbf{k})\cdot d\mathbf{k}, \qquad \mathbf{A}(\mathbf{k}) = i\langle u_{\mathbf{k}}|\nabla_{\mathbf{k}}|u_{\mathbf{k}}\rangle,
\end{equation}
where $\mathbf{A}(\mathbf{k})$ is the Berry connection, can take any value but are only well defined modulo $2\pi$ owing to the gauge freedom of the eigenvectors. However, when chiral symmetry confines $\mathbf{h}(k)$ to a plane ($h_z=0$) and the band is gapped, the phase of $h_x+ih_y$ accumulated around the loop is restricted to integer multiples of $2\pi$, associated with a full winding about the origin. In one-dimensional SSH-type lattices, this is evident from
\begin{equation}
\phi_z = \pi \mathcal{V} \ (\mathrm{mod} \ 2\pi), \qquad \mathcal{V} \in \mathbb{Z},
\label{Eq: Winding Quantisation}
\end{equation}
which relates the Zak phase to the winding number $\mathcal{V}$, highlighting the restriction of the accumulated phase to discrete values rather than a continuous range and giving a quantised topological classification.

\subsection{Chirality and Beyond-nearest neighbours \label{App: Chirality and BNN}}

The topological behaviour of the SSH model arises from its preserved symmetries, most notably chiral symmetry, which constrains the form of the Hamiltonian and protects the quantised topological phase set by the relative arrangement and strength of the spring connections. In the Bloch picture, this restricts the Hamiltonian to off-diagonal terms coupling dissimilar sublattices. Extending the SSH framework to include non-local couplings, the diatomic basis gives rise to two classes of beyond-nearest-neighbour (BNN) interactions: even-order connections between sites on the same sublattice and odd-order connections between sites on different sublattices, each giving rise to distinct topological responses. 

Even-order BNN couplings inherently break chiral symmetry, as evident from the dynamical matrix of the infinitely periodic system, where these additional interactions introduce asymmetric contributions to the diagonal elements. The dynamic behaviour for a diatomic lattice with a single even-order BNN connection is governed by the following equations of motion,
\begin{align}
    \begin{split}
        m\ddot{U}_n=&C_{\alpha,1} (Y_n-U_n)+C_{\beta,1}(Y_{n-1}-U_n)\\&+C_{\alpha,p}\left(U_{n+\frac{p}{2}}-U_n\right)+C_{\alpha,p}\left(U_{n-\frac{p}{2}}-U_n\right),\\[8pt]
        m\ddot{Y}_n =& C_{\beta,1}(U_{n+1}-Y_n)+C_{\alpha,1}(U_n-Y_n)\\&+C_{\beta,p}\left(Y_{n+\frac{p}{2}}-Y_n\right)+C_{\beta,p}\left(Y_{n-\frac{p}{2}}-Y_n\right),
        \label{Eq: Similar Sub-lattice BNN EoM}
    \end{split}
\end{align}
with the spring stiffnesses denoted by $(C_{\alpha,p})$ or $(C_{\beta,p})$, where the subscript $(p)$ specifies the order of the connection. Consistent with the SSH framework, alternating strong and weak NN ($p=1$) connections are always included, while the spatial extent of the BNN interactions ($p>1$) can assume any value satisfying $p \in 2\mathbb{Z}$. Likewise the variables $U_n$ and $Y_n$ denote the mass displacements within the $n^\mathrm{th}$ unit cell. 

Satisfaction of Bloch’s theorem, together with the assumption of harmonic displacement solutions of the form $\exp(i(kx - \omega t))$, allows the governing equations of motion to be generalised to a frequency-domain eigenvalue problem,
\begin{widetext}
\begin{equation}
    \begin{bmatrix}
        (C_{\alpha,1}+C_{\beta,1}) + 2C_{\alpha, p}(\cos(kp)-1)-m\omega^2  & -(C_{\alpha,1}+C_{\beta,1}e^{-2ik}) \\
        -(C_{\alpha,1}+C_{\beta,1}e^{2ik}) & (C_{\alpha,1}+C_{\beta,1}) + 2C_{\beta, p}(\cos(kp)-1)-m\omega^2
    \end{bmatrix}
    \begin{bmatrix}
        U_n \\
         Y_n 
    \end{bmatrix}=0.
    \label{Eq: Similar Sub-lattice Dynamic Matrix}
\end{equation}
\end{widetext}
Owing to the presence of non-uniform on-diagonal terms, assuming $C_{\alpha,p}\neq C_{\beta,p}$, it is not possible to reformulate the system into a purely off-diagonal Bloch Hamiltonian that satisfies the chiral anti-commutation relation. Consequently, chiral symmetry is broken, and the associated topological invariants are no longer quantised, preventing a well-defined topological classification framework.

Conversely, odd-ordered BNN connections preserve chiral symmetry, with the on diagonal terms remaining invariant to changes in the wavevector $(k)$, being determined solely by the negative sum of the spring stiffnesses. Thus the dynamic matrix \eqref{Eq: Reduced BNN Dynamic Matrix} can be reformulated to reveal the chiral symmetric Bloch Hamiltonian, 
\begin{equation}
    \underbrace{\left[ \Re(\gamma) \boldsymbol{\sigma}_x-\Im(\gamma)\boldsymbol{\sigma}_y\right]}_{\boldsymbol{\hat{H}}}\boldsymbol{u}+\left[\left(\eta-m\omega^2\right)\boldsymbol{\mathbb{I}}\right]\boldsymbol{u}=0.
\label{Eq: BNN Dynamic Matrix - Hamiltonian Reformulation append}
\end{equation}
in terms of the identity and Pauli matrices.

\section{Non-Locality and the Winding Number \label{app: winding bounds}}

The winding number provides a more complete classification of non local lattice topology. The Zak phase is a gauge dependent geometric phase describing the accumulation of Berry phase by occupied Bloch states across the Brillouin zone and depends on the choice of unit cell. In contrast, the winding number is defined from the dynamical matrix counting how many times its complex coupling structure or trajectory, $(\gamma)$ encircles the origin as the Brillouin zone is traversed,
\begin{align}
\begin{split}
    \mathcal{V} = \frac{1}{2\pi} \oint_{\mathrm{BZ}} d\theta
    &= \frac{1}{2\pi} \oint_{\mathrm{BZ}} \frac{\partial}{\partial k} \left( \Im \ln \gamma(k) \right) dk
    \\[5pt]&= \frac{1}{2\pi i} \oint_{\mathrm{BZ}} \frac{\gamma'(k)}{\gamma(k)} dk.
    \label{Eq: Winding Number append}
\end{split}
\end{align}

The origin is a singular point in chiral systems, where the phase of the off-diagonal Bloch function is undefined due to vanishing magnitude. At the point that the band gap closes the complex trajectory intersects the origin rather than encircling it. Thus, for the winding number to be well defined, $\gamma \neq 0 ~\forall~k$. Tuning through $\gamma = 0$ corresponds to a topological phase transition between regimes with distinct winding numbers.

The Winding number exposes the richer topology of non-local lattices that is not accessible through the Zak phase, revealing a heightened number of topological phase transitions, as well as exposing additional structure in the curvature of the complex trajectory contours. The added complexity of BNN interactions modifies the analytic structure of the dynamical matrix across the Brillouin zone, introducing higher-index exponential terms in the off-diagonal complex coupling $\gamma$. This increased polynomial index consequently facilitates higher-order windings beyond those present in nearest-neighbour models. The sign of the winding numbers governs whether the complex trajectory encircles the origin in a clockwise ($\mathcal{V}<0$) or anticlockwise ($\mathcal{V}>0$) direction.

\subsubsection{Winding Number Bounds \label{Sec: winding bounds}}

For each lattice structure, the spatial extent of its couplings imposes a bound on the attainable winding numbers, limiting the maximum topological order that can be supported. In the complex-plane formulation, this constraint is naturally expressed through contour integration methods based on residue calculus and the argument principle. Within this framework, the residue theorem states that contour integrals of analytic functions are determined entirely by the residues of the singularities enclosed by the contour.

Let
\begin{equation}
    f(z)=\sum_i \mu_i z^i
    = \mu_0 \prod_j (z-\zeta_j)
    \label{Eq: Complex Polynomial append}
\end{equation}
be some generalised meromorphic complex polynomial that exists within some contour $\mathcal{C}$. With $f(z)$ being analytic everywhere except at isolated poles, where it diverges to infinity in a well-defined manner, and $\mathcal{C}$ being a simple contour without intersections. The Argument principle states that the winding of $f(z)$ is equal to the net difference in the number of zeros $(\zeta)$ and poles $(\rho)$ inside the contour $\mathcal{C}$ counting multiplicity,
\begin{equation}
    \mathcal{V}=\frac{1}{2\pi i} \oint_{\mathcal{C}} \frac{f'(z)}{f(z)} dz =\sum_a n(\mathcal{C},\zeta_a)-\sum_b(\mathcal{C},\rho_b).
    \label{Eq: Argument Principle append}
\end{equation}

The Argument Principle is applied to the complex trajectory $\gamma$ of the non-local lattice structures to establish bounds on the admissible winding numbers. While $\gamma$ is initially expressed as a function of the wavenumber $k$, the analysis of its singularities within a simple closed contour is reformulated via a parametrisation of the Brillouin zone in terms of the complex contour variable $z$. Adopting the standard anti-clockwise convention, the contour is restricted to the unit circle $|z| = 1$, defined through the block-invariant mapping $z = \exp(i k l)$, where $l$ denotes the lattice spacing. Under this mapping, the trajectory is equivalently expressed as
\begin{equation}
    \gamma(z)=C_{\alpha,1} + C_{\beta,1} z^{-1} + C_{\alpha,p} z^{\frac{(p-1)}{2}} + C_{\beta,p} z^{-\frac{(p+1)}{2}},
    \label{Eq: Complex trajectory (mapped) append}
\end{equation}
where $l=2$. The expression is multiplied by $z^{\frac{p+1}{2}}$, producing an ordinary complex polynomial in $z$, 
\begin{equation}
    z^{\frac{(p+1)}{2}}\gamma(z)=C_{\alpha,1}z^{\frac{(p+1)}{2}} + C_{\beta,1}z^{\frac{(p-1)}{2}} + C_{\alpha,p} z^{p} + C_{\beta,p}.
    \label{Eq: Ordinary Complex trajectory (mapped) append}
\end{equation}
which can be evaluated using standard methods of complex analysis.

Substituting \eqref{Eq: Ordinary Complex trajectory (mapped) append} into the contour integral \eqref{Eq: Winding Number append} yields a formulation of the Argument Principle for the complex trajectory $\gamma(z)$,
\begin{widetext}
\begin{equation}
    \mathcal{V} = \frac{1}{2\pi i} \oint_{\mathcal{C}} \frac{\left(z^{\frac{(p+1)}{2}}\gamma(z)\right)'}{z^{\frac{(p+1)}{2}}\gamma(z)} dz=
    \frac{1}{2\pi i} \oint_{\mathcal{C}} \frac{\left(z^{\frac{(p+1)}{2}}\gamma(z)'\right)+\left(\frac{(p+1)}{2}z^{\frac{(p-1)}{2}}\gamma(z)\right)}{z^{\frac{(p+1)}{2}}\gamma(z)} dz,
\end{equation}
\end{widetext}
which, upon simplification, separates naturally into contributions from the zeros and poles of $\gamma(z)$. This is elucidated when written as
\begin{align}
\begin{split}
    \mathcal{V} = \frac{1}{2\pi i} \oint_{\mathcal{C}} \frac{\gamma(z)'}{\gamma(z)}&= \underbrace{\frac{1}{2\pi i} \oint_{\mathcal{C}} \frac{\left(z^{\frac{(p+1)}{2}}\gamma(z)\right)'}{z^{\frac{(p+1)}{2}}\gamma(z)} dz}_{\mathrm{Zeros}}\\[8pt]&-\underbrace{\frac{1}{2\pi i} \oint_{\mathcal{C}}\frac{(p+1)}{2}\frac{1}{z} dz}_{\mathrm{Poles}}.
\end{split}
\end{align}
where the contour integral of $\gamma(z)$ is recovered. There exists a single pole at $z=0$ with residue $\frac{p+1}{2}$, while the zeros are determined by the degree of the transformed function $z^{\frac{p+1}{2}}\gamma(z)$. The polynomial structure of \eqref{Eq: Ordinary Complex trajectory (mapped) append} spans powers from $z^{0}$ to $z^{p}$, and therefore admits up to $p$ zeros when counted with multiplicity, with the exact number depending on the specific lattice coefficients, and potentially reducing, under degeneracies, to fewer zeros including the limiting case of none. Therefore the span of admissible winding numbers ranges between,
\begin{align}
\begin{split}
    \mathcal{V}_{\max} = p - \frac{p+1}{2} = \frac{p-1}{2},\\[5pt]
    \mathcal{V}_{\min} = 0 - \frac{p+1}{2} = -\frac{p+1}{2}.
    \label{Eq: Winding Bounds append}
\end{split}
\end{align}

Despite the admissible range of winding numbers, not all values within this interval are necessarily attainable owing to the restricted orders of spatial interaction. With only nearest-neighbour and a single order of BNN couplings present, the resulting complex polynomial contains a sparse distribution of powers, particularly as the range of BNN interactions increases. Consequently, certain intermediate winding states cannot be realised. This is evident in Figure \ref{Fig: Dissimilar BNN Winding}d), where the state $|\mathcal{V}|=2$ is absent despite lying within the admissible winding range. The admissible winding interval is constrained by the lattice connection orders, yet the realised topological state is ultimately governed by the relative stiffnesses of the spring connections.

This is particularly transparent in the regime where a single order of spring connection dominates all other contributions. In this limit, Rouché’s theorem allows the zero-counting problem to be reduced to that of a simpler dominant function, with the number of zeros inside a closed contour determined solely by the leading-order term. Rouché’s theorem states that if two functions $g(z)$ and $r(z)$ are holomorphic within a region bounded by a contour $\mathcal{C}$ and satisfy $|g(z)|>|r(z)|$ on $\mathcal{C}$, then $g(z)$ and $g(z)+r(z)$ possess the same number of zeros within $\mathcal{C}$, counted with multiplicity. Applying this to $\gamma(z)$, when a single spring stiffness contribution exceeds the combined magnitude of all remaining interaction terms, the zero-counting problem reduces to that of the dominant contribution alone. For example, if the nearest-neighbour couplings satisfy $|C_{\alpha,1}|>|C_{\beta,1}|+|C_{\alpha,p}|+|C_{\beta,p}|$ or $|C_{\beta,1}|>|C_{\alpha,1}|+|C_{\alpha,p}|+|C_{\beta,p}|$, then the trajectory reduces to $z^{\frac{p+1}{2}}\gamma(z)\approx C_{\alpha,1}z^{\frac{p+1}{2}}$ and $z^{\frac{p+1}{2}}\gamma(z)\approx C_{\beta,1}z^{\frac{p-1}{2}}$, respectively. Whereby the resulting winding numbers reduce to the nearest-neighbour limits $\mathcal{V}=\{0,-1\}$, consistent with conventional local lattice systems.


%

\end{document}